\documentclass[a4paper]{article}
\usepackage{hyperref}
\usepackage{INTERSPEECH2019}
\usepackage{amsmath}
\usepackage{indentfirst}
\title{Probabilistic Permutation Invariant Training for Speech Separation}
\name{Midia Yousefi, Soheil Khorram, John H.L. Hansen}
\address{
	Center for Robust Speech Systems(CRSS)\\ The University of Texas at Dallas, Richardson, Texas, USA}
\email{\{Midia.Yousefi, Soheil.Khorram, John.Hansen\}@utdallas.edu}

\begin{document}
	
	\maketitle
	\begin{abstract}
		Single-microphone, speaker-independent speech separation is normally performed through two steps: \emph{(i)} separating the specific speech sources, and \emph{(ii)} determining the best output-label assignment to find the separation error. The second step is the main obstacle in training neural networks for speech separation. Recently proposed \emph{Permutation Invariant Training (PIT)} addresses this problem by determining the output-label assignment which minimizes the separation error. In this study, we show that a major drawback of this technique is the overconfident choice of the output-label assignment, especially in the initial steps of training when the network generates unreliable outputs. To solve this problem, we propose \emph{Probabilistic PIT (Prob-PIT)} which considers the output-label permutation as a discrete latent random variable with a uniform prior distribution. Prob-PIT defines a log-likelihood function based on the prior distributions and the separation errors of all permutations; it trains the speech separation networks by maximizing the log-likelihood function. Prob-PIT can be easily implemented by replacing the minimum function of PIT with a soft-minimum function. We evaluate our approach for speech separation on both TIMIT and CHiME datasets. The results show that the proposed method significantly outperforms PIT in terms of Signal to Distortion Ratio and Signal to Interference Ratio.

	\end{abstract}
	\vspace{-0.05in}
	\section{Introduction}
	\noindent
	Humans are equipped with effective abilities to efficiently focus on a particular sound received through their auditory system~\cite{assmann2004perception}. In the cocktail party scenario, humans are able to isolate a target speech signal from a mixture of conversations with little effort~\cite{bregman1994auditory, darwin1995auditory}. However, speech separation is still a challenging task for machines~\cite{divenyi2004speech}, and there has been continuing research on how the human auditory system manages to separate different sound sources. 
	
	The goal of speech separation is to separate a set of speech signals from a set of mixed signals. In this paper, we focus on single-microphone, speaker-independent source separation, which has many potential applications such as: digital hearing aids~\cite{healy2017algorithm, van2017eeg}, automatic speech recognition~\cite{erdogan2017deep, rennie2010single, weng2015deep}, speaker diarization~\cite{sell2017multi, von2019all}, emotion recognition~\cite{khorram2019jointly, gideon2017progressive}, speaker verification and identification~\cite{khalil2019robust, hansen2015speaker}. 
	Researchers have proposed many methods to solve the speech separation task; these methods can be divided into two categories: (1) \emph{unsupervised}, and (2) \emph{supervised}.
	
	\textbf{Unsupervised} --
	Computational Auditory Scene Analysis (CASA) is one of the first unsupervised attempts to address speech separation~\cite{ref1}. CASA is a knowledge-based system which uses substantial knowledge regarding the human auditory system. Another popular unsupervised approach is Independent Component Analysis (ICA)~\cite{comon1994independent} which decomposes the mixture signal into a linear combination of statistically independent sources. These unsupervised approaches can be used in real-time processing, but it is difficult to incorporate detailed statistical knowledge into these approaches~\cite{roweis2001one}. 
	
	\textbf{Supervised} --
	These methods exploit generative modeling techniques to solve the separation task. They first build speaker-dependent models for each speaker in the training database and then use these models to perform the separation task. One popular model-based method is Non-negative Matrix Factorization (NMF) \cite{smaragdis2007convolutive, yousefi2016supervised} which represents the speech signal as a linear combination of its hidden structure. The linearity of NMF is a major drawback, which prevents NMF to capture the complex structure of speech.
	
	Lately, there has been increasing interest in nonlinear models, specifically, Deep Neural Networks (DNNs) \cite{wang2014training, huang2015joint, zhang2016deep, Mamun2019convolutional}. In Deep Clustering (DPCL) \cite{hershey2016deep, isik2016single}, first, the time-frequency bins of the mixtures are mapped into an embedding space; then, a clustering algorithm is performed in the embedding space; finally, a binary mask is generated based on each cluster to reconstruct speech of each speaker. This approach minimizes the error in the embedding space, which does not necessarily lead to the minimization of the reconstruction error. In Deep Attractor Network (DANet)~\cite{chen2017deep}, cluster centroids called attractor points are created. Then, the time-frequency bins of each speaker are classified to these clusters.
	
	Recently proposed Permutation Invariant Training (PIT)~\cite{yu2017permutation, kolbaek2017multitalker} trains a neural network that separates the speaker-specific speech signals, and then determines the best output-label assignment which minimizes the separation error. Finding the best output-label assignment has been a challenge in speech separation, which is referred to as label permutation ambiguity. PIT employs a hard decision to choose the output-label assignment. This approach is suboptimal, especially in the initial steps of training when the network generates unreliable outputs and the costs of different permutations are close. In this paper, we show that updating network parameters based on the cost of one single permutation is not an optimal solution and leads to an inefficient training of the network.
	
	In this study, we propose a new method called Probabilistic PIT (Prob-PIT) which considers the output-label permutation as a discrete latent random variable with a uniform prior distribution. Prob-PIT defines a log-likelihood function based on the prior distributions and the separation errors of all possible permutations. Next, the network is trained by maximizing the log-likelihood function. Unlike the conventional PIT that uses one output-label permutation with the minimum cost, Prob-PIT uses all permutations by employing the soft-minimum function. To show the effectiveness of the Prob-PIT, we first perform preliminary experiments on the TIMIT dataset to study the disadvantages of the hard output-label assignment performed in PIT. Then, we compare the Prob-PIT with the PIT on the TIMIT and CHiME datasets. The results demonstrate that Prob-PIT significantly outperforms PIT in terms of both (Signal to Distortion Ratio) SDR and (Signal to Interference Ratio) SIR. 
	\vspace{-0.1in}
	\section{Problem setup and preliminary experiment}
	\noindent
	In single-channel speech separation, we assume that the speech signals have been linearly mixed: $y[n]=\sum_{s=1}^{S} x_s[n]$, where $S$ is the number of sources; the goal is to extract all speech signals (i.e., $\{x_s\}_{s=1}^{S}$) from the mixed-signal (i.e., $y$). 
	To do so, signals are normally transferred to the frequency-domain using the Short-Time Fourier Transform (STFT), since STFT provides a better representation for harmonics, formants, and energy densities. Estimating phase information of this STFT representation is still an open problem \cite{williamson2016complex}. In most speech separation systems,
	phase information is borrowed from the mixed signal, which simplifies the speech separation to the task of estimating the magnitude spectra of the speech signals. Model-based approaches have gained increasing attention in recent years. However, due to the label permutation ambiguity problem, training a data-dependent model for speech separation is challenging. 
	
	\textbf{Label permutation ambiguity} -- Assume we are given a mixture signal, $y=x_1+x_2$, and the goal is to extract $x_1$ and $x_2$ from this mixture. We can employ a network with two outputs, $o_1$ and $o_2$, to solve this problem. However, there are two solutions to this problem: (1) $o_1=x_1$, $o_2=x_2$, and (2) $o_1=x_2$, $o_2=x_1$. Generally speaking, for $S$ sources in the mixture, there are $S!$ different solutions (permutations), which causes $S!$ different cost functions. In neural network training, we need to find the correct solution (correct cost function) and then perform the back-propagation algorithm through the correct cost. PIT performs a hard decision on choosing the best solution (solution with the minimum cost). This is not an efficient approach, especially in the initial epochs where the costs of different permutations are comparable. Forcing the network to be updated based on the minimum cost in PIT causes suboptimal training. To visualize this problem, we conduct a preliminary experiment on the TIMIT dataset, which we perform two-talker speech separation. For generating the mixtures, different speakers are chosen randomly. The SIR of the speakers is chosen randomly between 0 to 5dB with a uniform distribution.
	
\begin{figure}[t!]
	    \centering
	    \includegraphics[width=0.8\linewidth]{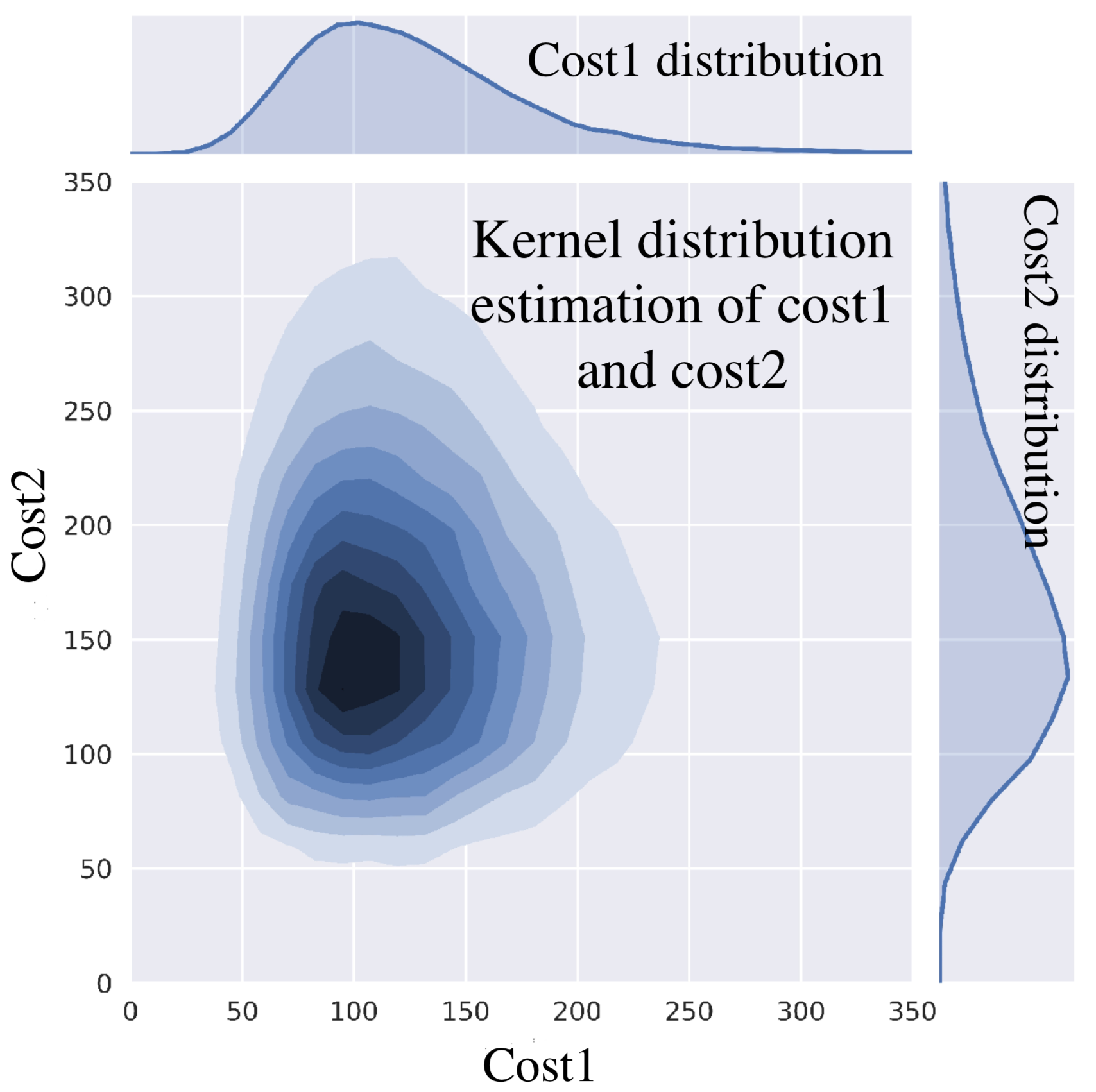}
	    \caption{Preliminary experiment on the TIMIT dataset for speech separation of a dual-speaker mixed signal using PIT. Cost1 and cost2 are the separation errors for the two possible output-target assignments. Cost1 and cost2 are highly likely to have close values in the first epoch.\vspace{-0.4cm}}
	    \label{fig:cost}
	\end{figure}

	\textbf{Preliminary experiment} -- We train a 2-layer LSTM neural network. The input is a sequence of 129-dim STFT magnitude spectral features computed over a Hamming window with a frame size of 32 ms and a 16 ms frame shift. The network has two outputs for estimating two 129 $\times$ $M$ streams of separated speech signals where $M$ is the number of frames. The Minimum Mean Square Error (MMSE) is used as the cost function and the network is trained based on the target-label assignment that gives the minimum cost among all permutations (PIT assumption). Fig. \ref{fig:cost} depicts the Kernel Distribution Estimation (KDE) of cost1 and cost2 for each data sample used in the first epoch. KDE gives us a sense on how cost1 and cost2 are distributed. As shown in Fig.\ref{fig:cost}, in the first epoch, cost1 and cost2 are more likely to be observed in regions where their values are very close. Therefore, choosing the minimum cost may lead to assigning the wrong target label to the network output. This problem affects the quality of the trained model and is more probable in the initial iterations of training where the network is still naive.
	
	\section{Probabilistic Permutation Invariant Training (Prob-PIT)}
	\noindent
	This section explains the details of the proposed Prob-PIT modeling technique. Assume $X_s$ is a high-dimensional vector containing the magnitude spectra of the $s$-th source; $X=[X_1, \dots, X_S]$ is a sequence of all $X_s$ vectors; and $Y$ contains the magnitude spectra of the mixed signal. In a model-based speech separation, we are given pairs of $(Y, X)$ and the goal is to find a model, that takes $Y$ and generates $X$. 
	
\begin{figure}[t]
		\centering
		\includegraphics[width=8.3cm]{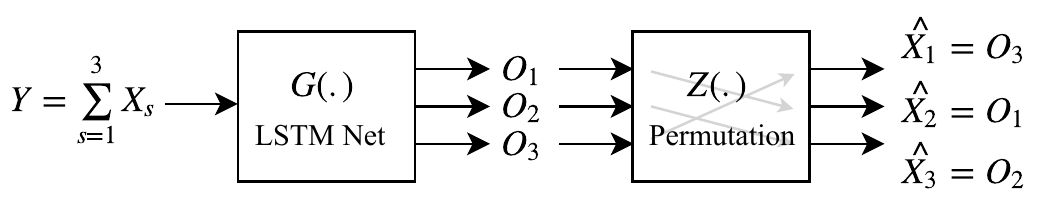}
		\vspace{-0.2in}
		\caption{Speech separation block diagram. $G(.)$ is a LSTM-based model and $Z(.)$ is a one-to-one permutation function.}
		\vspace{-0.1in}
		\label{fig:1}
	\end{figure}

	\textbf{Model structure} -- In this paper, we propose a new generative model to solve the speech separation task. The model is shown in Fig.~\ref{fig:1} and it can be expressed by:
	\vspace{-0.05in}
	\begin{equation}
	X=Z(G(Y,\theta)) + \epsilon,
	\label{eq:graphical_model}
	\vspace{-0.05in}
	\end{equation}
	where $G(.)$ is a neural network with the learnable parameters of $\theta$; $O=[O_1, \dots,O_S]$ is the output of the separator network (i.e., $O=G(Y,\theta)$); $Z(.)$ is a one-to-one permutation function; $\hat{X}=[\hat{X}_1, \dots,\hat{X}_S]$ is the output of the permutation function (i.e., $\hat{X}=Z(O)$); and $\epsilon$ is the estimation error. 
	
	The network, $G(.)$, takes the mixed signal, $Y$, and performs the speech separation by estimating speech signals; however, the order of the generated signals, $O$, may not be consistent with the order of the target signals, $X$. Therefore, to calculate the separation error we need a permutation function. The permutation function, $Z(.)$, permutes the order of $O$ to match the order of the target signals $X$.
	
	The model proposed in Eq.~(\ref{eq:graphical_model}) has two latent random components: $\epsilon$ and $Z(.)$. $\epsilon$ is the estimation error which is normally modeled by a standard Gaussian distribution with mean zero and variance $\sigma^2$. In this paper, we assume $\sigma$ is a hyper-parameter of the network and we tune it during the training phase. The permutation function, $Z(.)$, can take $S!$ different forms. We assume all these $S!$ forms are possible and all have the same probability of $\frac{1}{S!}$; $Z(.)$ follows the uniform distribution. For example, assume we are given two sources, then the function $Z(.)$ can take two forms: $z_1(.)$ and $z_2(.)$ such that $[a,b]=z_1([a,b])$ and $[b,a]=z_2([a,b])$.

	\textbf{Model training} -- We leverage the maximum log-likelihood method to train the proposed model. To do so, we first drive the log-likelihood expression of the model. According to the model explained in Eq.~(\ref{eq:graphical_model}) and considering the Gaussian distribution of $\epsilon$, the  probability of a clean speech, $X$, given a mixture signal, $Y$, and a specific permutation $Z$ can be written as:
	\begin{equation}
	P(X|Z,Y)=\mathcal{N}(Z(G(Y,\theta)),\,\sigma^{2}I),
	\label{eqn:prob}
	\end{equation}
	where, $I$ is the identity matrix and $\mathcal{N}$ is the Gaussian distribution. On the other hand, Bayes rule offers the following Equation to calculate the likelihood of $X$ given $Y$:
	\vspace{-0.05in}
	\begin{equation}
	P(X|Y)=\sum_{All\ possible\ Z} P(X|Z,Y)P(Z),
	\label{eqn:prob1}
	\vspace{-0.05in}
	\end{equation}
	where the summation is taken over all possible permutations. $P(Z)$ is the prior distribution of a specific permutation. Due to the uniform prior assumption, $P(Z)$ is independent of $Z$ and is equal to $\frac{1}{S!}$. Considering the prior distribution of $P(Z)$ and substituting Eq.~(\ref{eqn:prob}) in Eq.~(\ref{eqn:prob1}), the following expression is obtained for the log-likelihood function $Q = \log P(X|Y)$:
	\vspace{-0.05in}
	\begin{equation} \label{eqn:expanded_cost}
	\begin{split}
	Q(\theta) = C + \log \sum_{All\ Z}
	exp{(\dfrac{-||X-Z(G(Y,\theta))||^2}{\gamma})},
	\end{split}
	\vspace{-0.2in}
	\end{equation}
	where $\gamma=2\sigma^2$, and $C$ is a constant value that does not depend on the learnable parameters $\theta$. It depends on the number of sources, number of frames, dimensionality of features and variance of the estimation error. We maximize the log-likelihood function, expressed by Eq.~(\ref{eqn:expanded_cost}), to train the model.
	
	To ensure numerical stability of the Eq.~(\ref{eqn:expanded_cost}), we employ the log-sum-exp stabilization trick: $\log \sum_i e^{x_i} = \max_i x_i + \log \sum_i e^{x_i-\max_i x_i}$. Following equations show the numerically stable form of the Eq.~(\ref{eqn:expanded_cost}):
	\begin{equation} \label{eqn:final}
	\begin{split}
	Q(\theta)=-g(Z_{min},\theta) + \hspace{4cm} \\
	\gamma \ \log\bigg( 1 + 
	\sum_{Z \neq Z_{min}} \exp\Big(
	\frac{g(Z_{min},\theta)-g(Z,\theta)}{\gamma}
	\Big)
	\bigg),\hspace{0.2cm} \\[0.1cm]
	g(Z,\theta)=||X-Z(G(Y,\theta))||^2, \hspace{2.8cm} \\[0.1cm]
	Z_{min} = \arg\min_{Z} g(Z,\theta).\hspace{3.7cm} 
	\end{split}
	\vspace{-0.07in}
	\end{equation}
	In this equation, $g(Z,\theta)$ is the separation error of the permutation $Z$ and $Z_{min}$ is the permutation that has the minimum separation error. As mentioned before, $\gamma$ is equal to $2\sigma^2$ and $\sigma^2$ is the variance of the estimation error in the proposed model. Since $g(Z_{min})-g(z)$ is always negative, both exponential and logarithmic functions are numerically stable.
	
	According to Eq.(\ref{eqn:final}), $Q(\theta)$ can be calculated by applying the smooth minimum~\cite{cuturi2017soft} of the costs of all permutations with a smoothing factor of $\gamma$. Conventional PIT uses the permutation that minimizes the mean square error of the clean speech and its corresponding separated signal. This is the same as minimizing $g(Z_{min})$. In other words, with $\gamma=0$, maximizing Eq.~(\ref{eqn:final}) is equal to minimizing the PIT cost function and therefore PIT is the same as Prob-PIT with $\gamma=0$.
	
	In contrast to PIT,  Prob-PIT considers the costs of all possible permutations to train the network. In  Eq.(\ref{eqn:final}), $\gamma$ has an important role. It provides a compromise between the cost of the minimum permutation and the cost of all permutations. A larger $\gamma$ results in paying more attention to all possible output-label permutations.
	
	\vspace{-0.1in}
	
	\begin{figure*}
		
		\centering
		
		\begin{tabular}{cccc}
			
			\vspace{-1.5mm}
			\hspace{-4mm}
			\includegraphics[width=42mm]{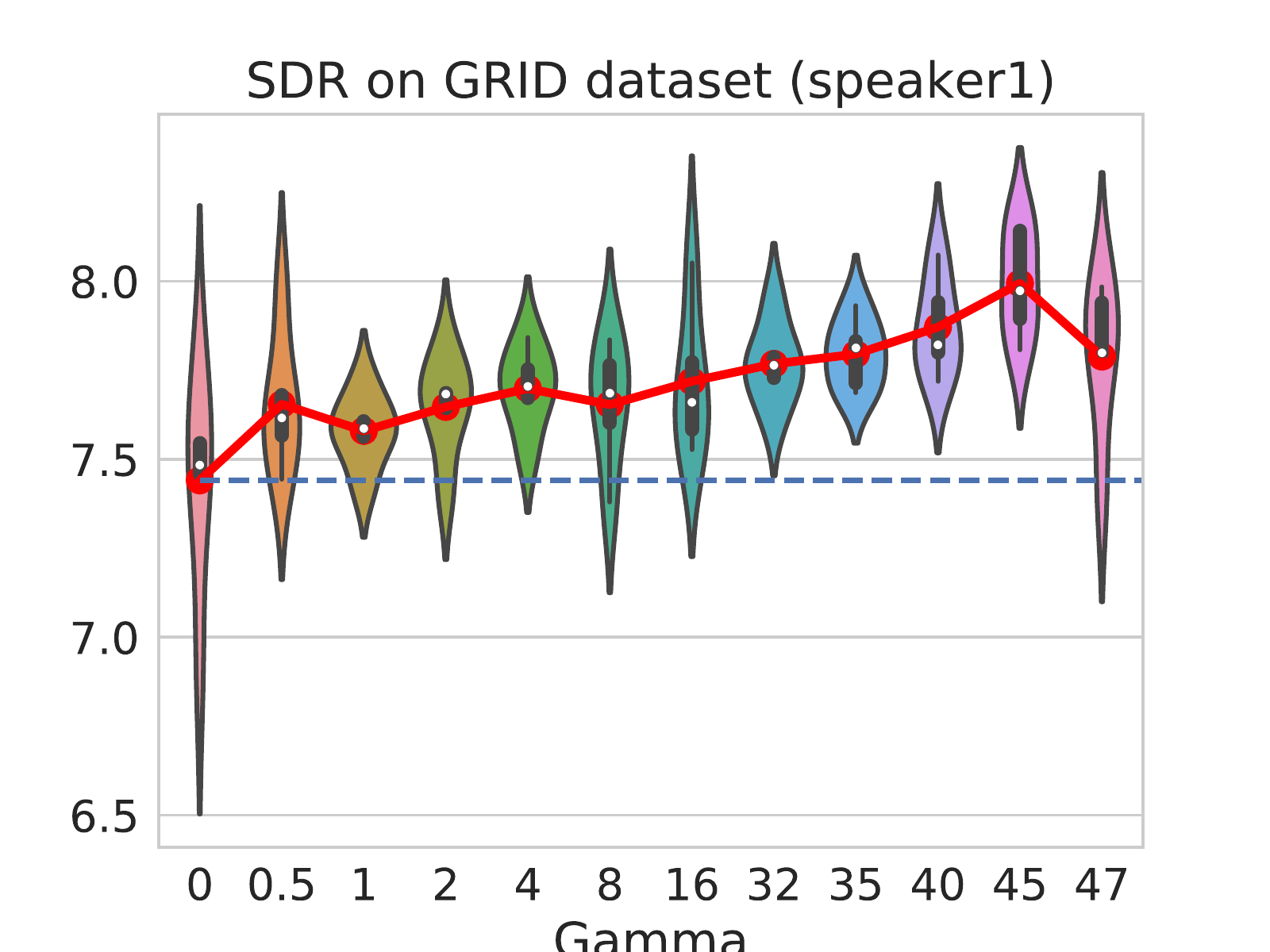}&
			\hspace{-4mm}
			\includegraphics[width=42mm]{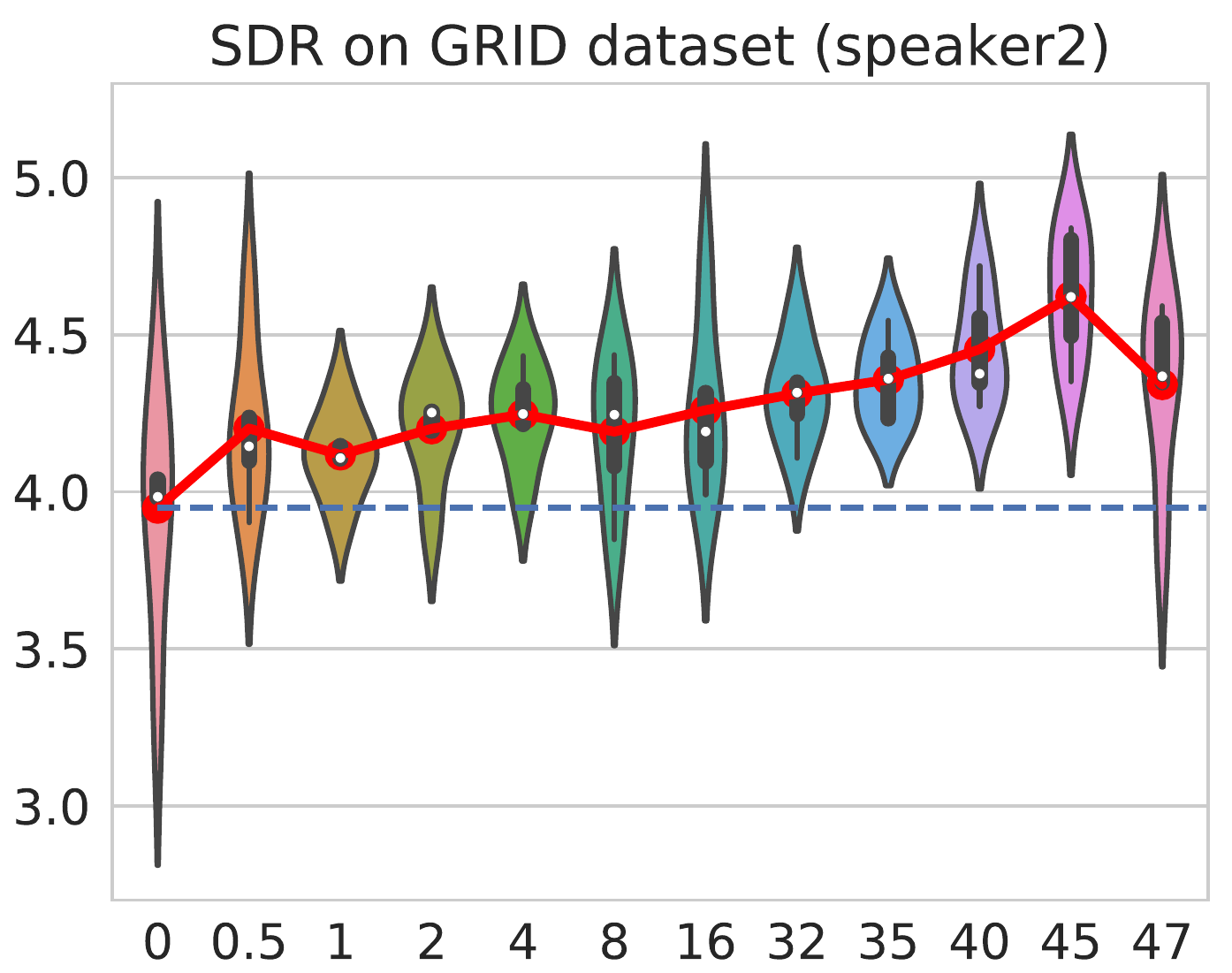}&
			\hspace{-4mm}
			\includegraphics[width=42mm]{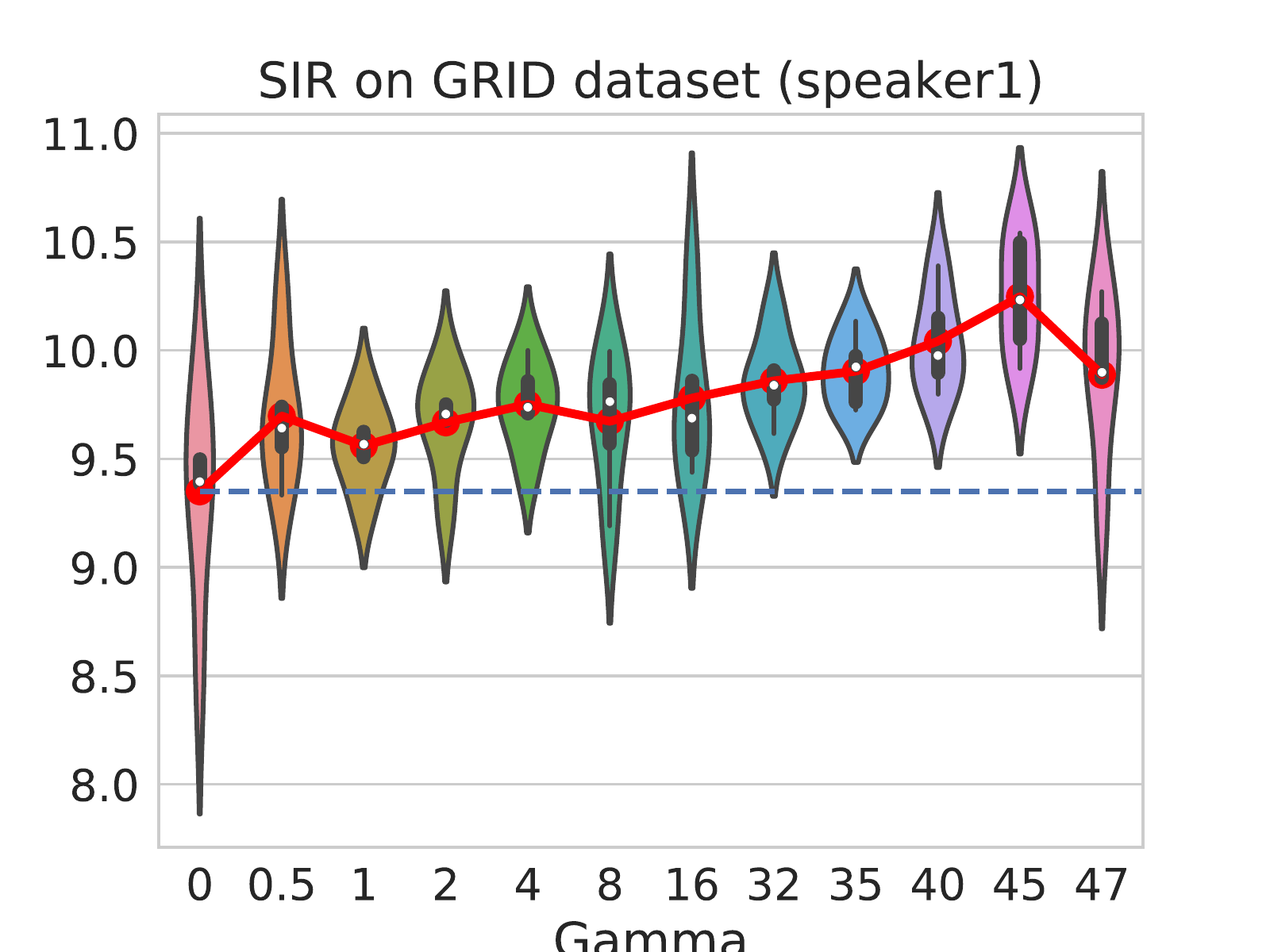}&
			\hspace{-4mm}
			\includegraphics[width=39mm]{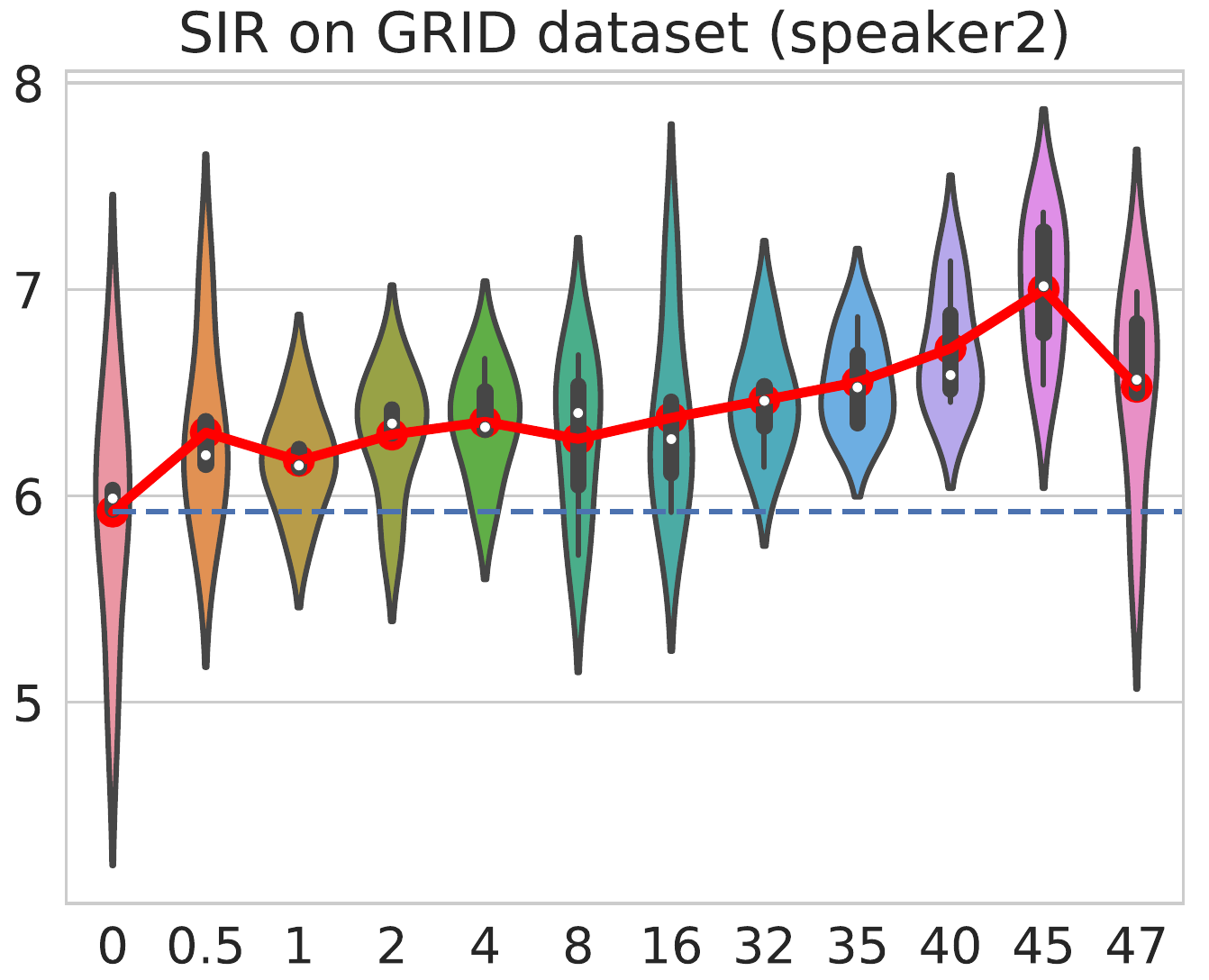}\\
			\footnotesize Gamma&\footnotesize Gamma&\footnotesize Gamma&\footnotesize Gamma\\
			\vspace{-1.5mm}
			\hspace{-4mm}
			\includegraphics[width=42mm]{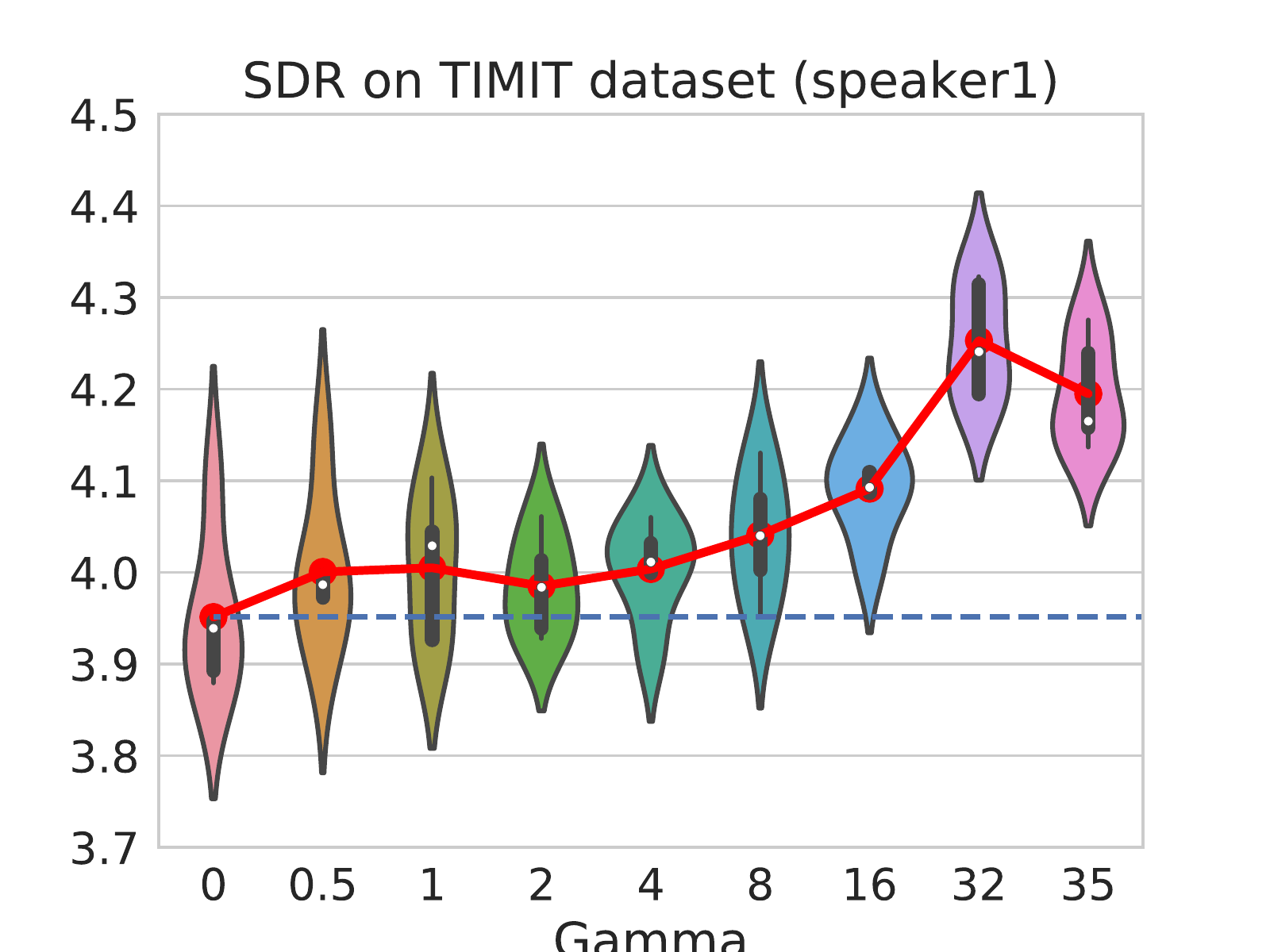}&
			\hspace{-4mm}
			\includegraphics[width=42mm]{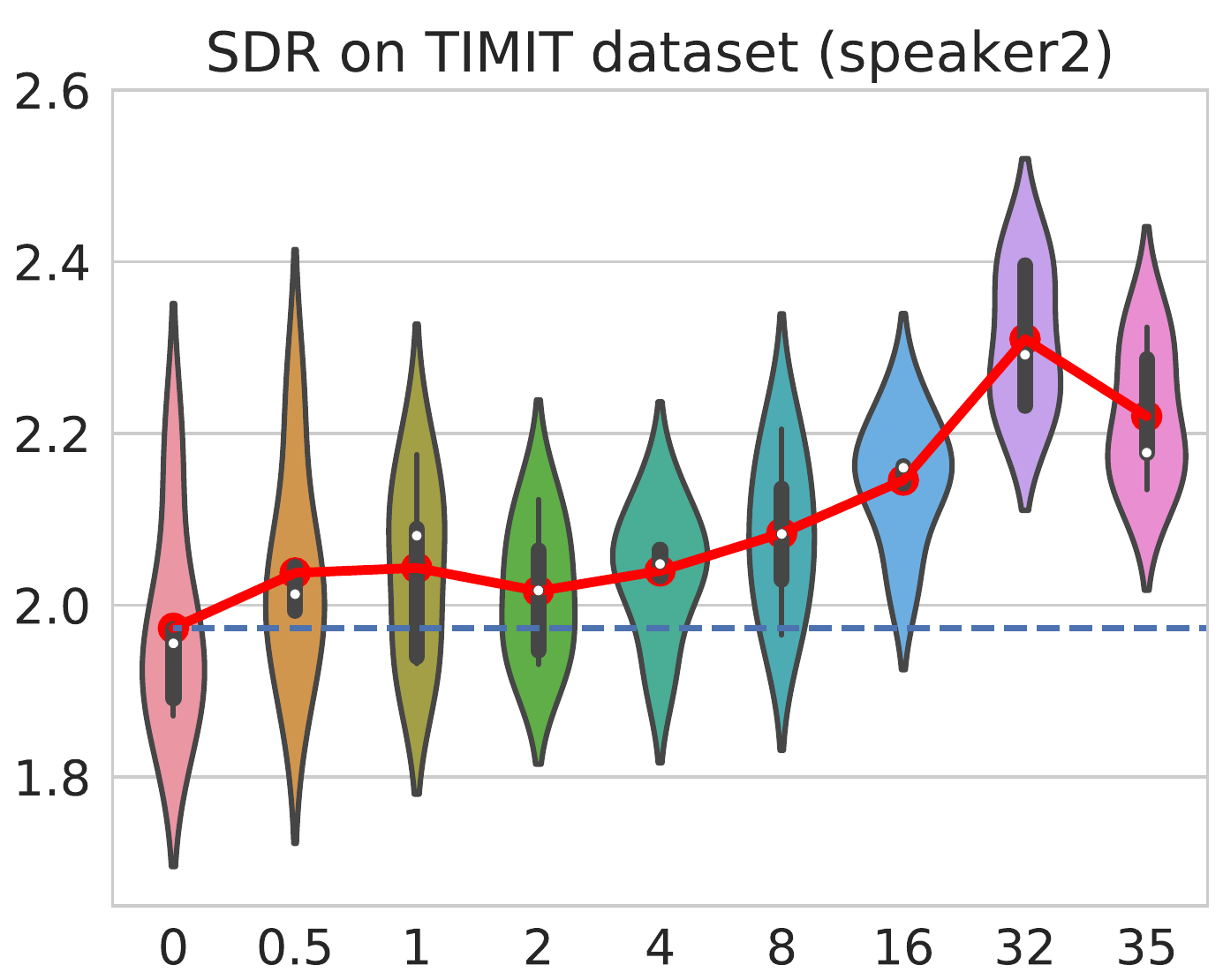}&
			\hspace{-4mm}
			\includegraphics[width=42mm]{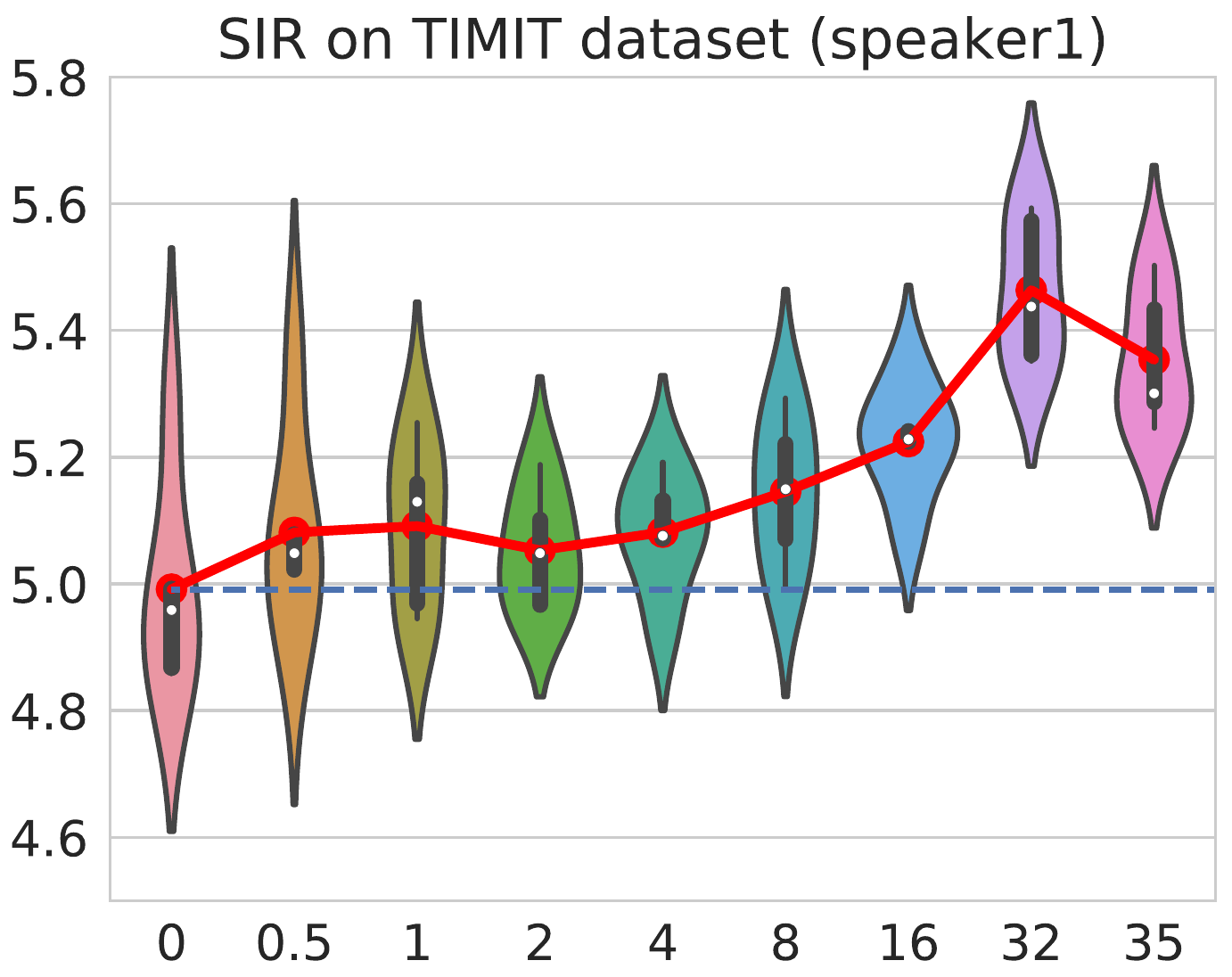}&
			\hspace{-4mm}
			\includegraphics[width=42mm]{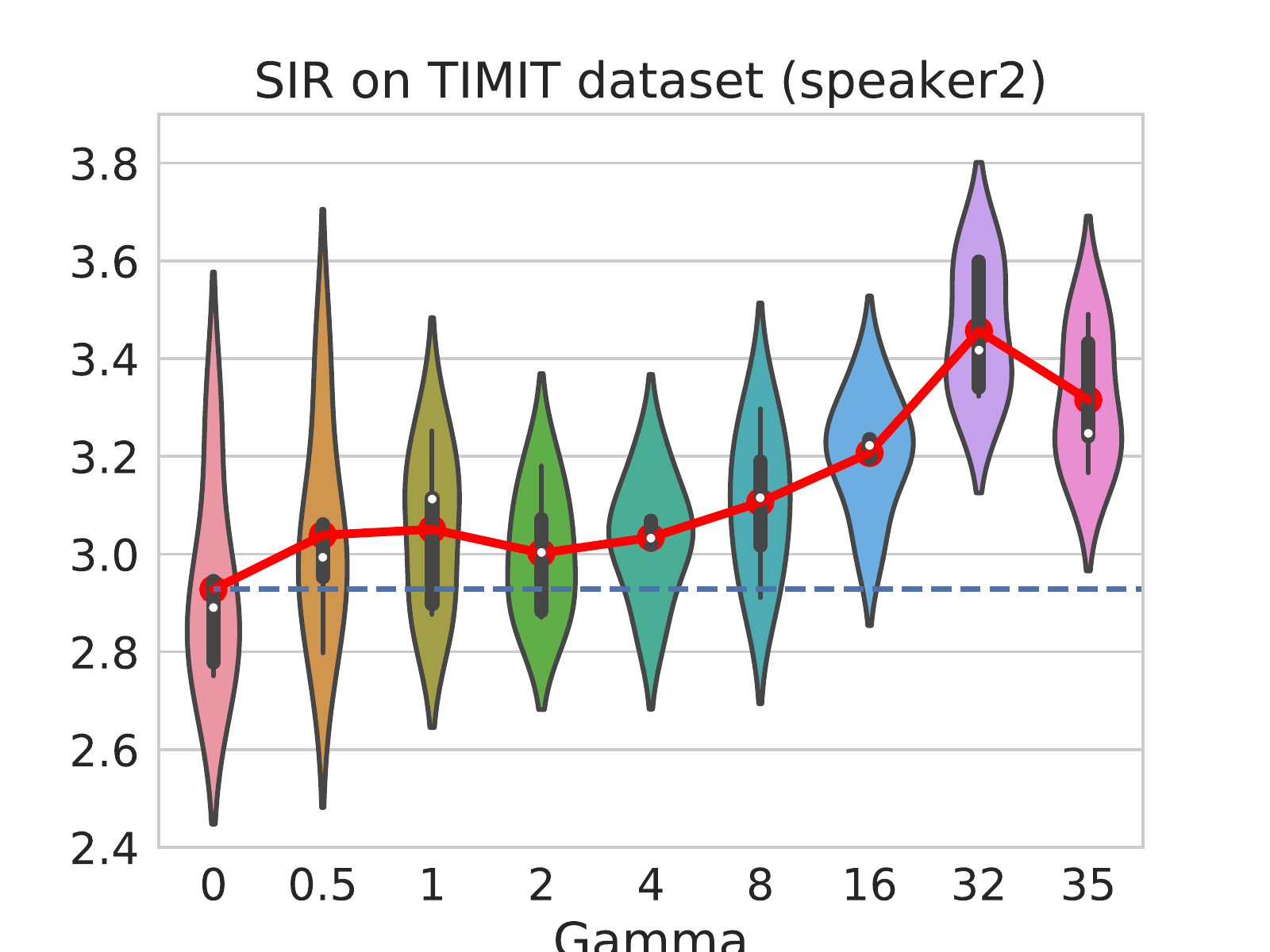}\\
			\footnotesize Gamma&\footnotesize Gamma&\footnotesize Gamma&\footnotesize Gamma\\
		\end{tabular}
		\vspace{-0.1in}
		\label{figur}\caption{Speech separation performance on the TIMIT and GRID datasets. For each $\gamma$, five experiments are performed (total of 105  experiments for both datasets). Each violin-shaped object represents the boxplot and the kernel distribution estimation of those five experiments. The blue dashed line, which is the mean of the results obtained for $\gamma=0$, shows the conventional PIT baseline. The red circles on the red solid line are the mean of the five experiments for different values of $\gamma$. Prob-PIT outperforms PIT for all values of $\gamma$. Also, the best separation performance is achieved by $\gamma=45$ for the GRID dataset and $\gamma=32$ for the TIMIT dataset.}
		\vspace{-0.2in}
		\label{fig:results}
	\end{figure*}
	
	\section{Experiments, results and discussion}
	\noindent
	\label{sec:exp}
	\textbf{Dataset} --
	To evaluate the effectiveness of Prob-PIT, several experiments are conducted on the TIMIT and GRID datasets. The GRID is a  multi-speaker, sentence corpus~\cite{cooke2006audio}, which has been used in monaural speech separation and recognition challenge~\cite{cooke2010monaural}. Additionally, this corpus has been widely used for assessing the perception of simultaneous speech signals~\cite{shokouhi2017teager,yousefi2018assessing,tu2015speech}. This corpus consists of 34 subjects (18 male and 16 female speakers), each narrating 1000 sentences.
	In this study, a 10h training set of mixed speech is prepared, which is composed of artificially summed random sentences from randomly chosen speakers. The SIR of the sentence pairs is uniformly distributed between 0 to 5dB. We have also generated 4h of mixed speech as the validation set and 2h as the test set. Also, using the TIMIT dataset, 8h of the training set, 3h of the validation and 1h of the test set is prepared with the same procedure we used for the GRID corpus. 
	
	\textbf{Performance evaluation} --
	To evaluate the performance of the proposed Prob-PIT, we employ the widely used blind source separation evaluation (BSS-EVAL) toolbox~\cite{vincent2006performance}. In speech separation, there are two types of ``noise'' in the separated signal. Noise due to the mis-separation which is called interference, and noise due to the reconstruction algorithm itself. SDR and SIR are good metrics to measure the amount of remained interference and reconstruction noise in the separated signal.
	SIR is defined as the ratio of the target signal power to that of the interference signal still remained in the separated speech. SDR is defined as the ratio of the target signal power to the distortion introduced by the interference and reconstruction noise.  We report SDR and SIR because they have been shown to be well correlated with human assessments of signal quality~\cite{fox2007modeling}.
	
	\textbf{Model} --
	In this study, we consider the case of two-talker mixed speech separation. The network architecture we use is one of the effective structures employed by conventional PIT in~\cite{kolbaek2017multitalker}, which consists of one feed-forward layer followed by two recurrent LSTM layers, 128 neurons each, and a softmax layer with two units that performs the separation task for two-talker speech signal. Since LSTM maintains speaker-specific information extracted from previous frames, it is quite suitable for the task of speaker-independent speech separation~\cite{chen2017long}. The network is trained for 50 epochs with a dropout rate of 20$\%$ and a learning rate of 0.0005 reduced by 0.7 when the cross-validation loss improvement is less than 0.003 in two successive epochs. The input of the network is a 129-dim STFT magnitude spectra computed over a frame size of 32ms with 50$\%$ of frame shift. In the output of the network, two 129 * $M$ streams of the magnitude spectra are generated for the two speech sources.
	\subsection{Experiments and discussions}
	\noindent
	We evaluated the proposed Probabilistic PIT on two-speaker speech separation problem. As mentioned in Sec.3, $\gamma$ is the smoothness parameter which converts PIT $(\gamma=0)$ to the Prob-PIT. For Probabilistic PIT, different values of $\gamma$ in the exponential range $(2^{-1},2^0,2^1...)$ are chosen until a decrease in the performance is observed. In all the experiments, for $\gamma \leqslant 32$ the separation performance increases with increasing $\gamma$. Therefore, to study the optimal value of $\gamma$, we take smaller step-size for $32<\gamma<64$. In order to minimize the effect of parameter initialization on our final separation metrics, we train each network five times with different initial parameters.
	
	The STFT magnitude of the mixed and clean speech signals are used to train the LSTM networks. In the training step, 50 epochs are completed with a batch size of 32 and a fixed $\gamma$. The network parameters are updated with respect to the gradients of Eq.(\ref{eqn:final}) using Adam optimization algorithm. During the test phase, the evaluation metrics, SDR and SIR are computed for the estimated speech waveforms. Fig.\ref{fig:results} shows the results of the experiments on both TIMIT and GRID datasets. In this figure, each violin-shaped object is the boxplot of the 5 experiments with their kernel distribution estimation. Each boxplot displays the minimum, the first quartile, the median, the third quartile and the maximum points of the results obtained from five experiments. The white points represent the median values. Additionally, the kernel distribution estimation depicts the probability distribution of the results. The more result points are in a specific range, the larger the violin is for that range. Also, each red circle on the red line represents the mean of the evaluation metrics in 5 experiments for each $\gamma$. The blue dashed line is the mean of the performance metrics for $\gamma=0$ as the PIT baseline.  
	
	The top four plots in Fig.\ref{fig:results}, shows the results of experiments on the GRID dataset. As can be seen in the figures, SDR and SIR for both speakers in all values of the $\gamma$ outperform the PIT baseline. The separation performance is at its maximum for $\gamma=45$ which is consistent for all the metrics on both speakers. The other four plots show the results on the TIMIT dataset. Again, the Prob-PIT for all values of $\gamma$ has a better performance in terms of SDR and SIR. $\gamma=32$ gives the best separation evaluation metric on this dataset. We also perform the pairwise t-test to evaluate the statistical significance of the Prob-PIT compared to the PIT ($\gamma=0$). The results of the t-test demonstrate that our proposed system is significantly (p-value $ < 0.01$) better than the PIT baseline for $1<\gamma$ for the GRID dataset and $8<\gamma<40$ for the TIMIT dataset.
	
	\textbf{Discussion} -- Several reasons explain the superiority of the proposed method over conventional PIT. \emph{First}, the conventional PIT applies a hard decision on assigning the output-label permutation that minimizes the total separation error. This is not an efficient decision especially in the initial steps of training when the network is unable to perform an effective separation. Hence, in the probabilistic PIT, we consider the costs of all possible permutations for training the network. \emph{Second}, the minimum cost function used in PIT is replaced by a soft-minimum function in Prob-PIT. In several applications of machine learning~\cite{cuturi2017soft}, it has been demonstrated that replacing the minimum by the soft-minimum results in a smoother optimization landscape and therefore it is less likely to converge to a poor local minimum. Our results confirm this finding for speech separation as well. Two observations confirm this finding: (1) SDR and SIR values of Prob-PIT are generally better than PIT; (2) the variance of the SDR and SIR values are lower for reasonable choices of Gamma ($1<\gamma<35$). A lower variance in the results shows a more stable system, which may be caused by a smoother optimization landscape.
	\vspace{-0.1in}
	
	\section{Conclusion}
	\noindent
	In this study, we proposed the probabilistic PIT to address the single-channel, speaker-independent speech separation. A long-lasting problem in speech separation task is finding the correct label for each separated speech signal, which referred to as \emph{label permutation ambiguity}. Recently proposed PIT solves this challenge by training a neural network based on the output-label assignment with minimum separation cost. The hard choice of the minimum cost permutation is not the best technique especially in the initial epochs of the training where the network is still not strong enough to effectively separate the speech signals. Contrary to PIT, in our proposed Prob-PIT, we consider all possible permutations as a discrete latent variable with a uniform prior distribution. Next, we train the network by maximizing the log-likelihood function defined based on the prior distributions and the separation errors of all possible permutations. The results of our proposed approach on the TIMIT and CHiME datasets show that the proposed Probabilistic PIT significantly outperforms PIT in terms of SDR and SIR.

	\bibliographystyle{IEEEtran}
	
	\bibliography{mybib}

\begin{thebibliography}{10}
\providecommand{\url}[1]{#1}
\csname url@samestyle\endcsname
\providecommand{\newblock}{\relax}
\providecommand{\bibinfo}[2]{#2}
\providecommand{\BIBentrySTDinterwordspacing}{\spaceskip=0pt\relax}
\providecommand{\BIBentryALTinterwordstretchfactor}{4}
\providecommand{\BIBentryALTinterwordspacing}{\spaceskip=\fontdimen2\font plus
\BIBentryALTinterwordstretchfactor\fontdimen3\font minus
  \fontdimen4\font\relax}
\providecommand{\BIBforeignlanguage}[2]{{%
\expandafter\ifx\csname l@#1\endcsname\relax
\typeout{** WARNING: IEEEtran.bst: No hyphenation pattern has been}%
\typeout{** loaded for the language `#1'. Using the pattern for}%
\typeout{** the default language instead.}%
\else
\language=\csname l@#1\endcsname
\fi
#2}}
\providecommand{\BIBdecl}{\relax}
\BIBdecl

\bibitem{assmann2004perception}
P.~Assmann and Q.~Summerfield, ``The perception of speech under adverse
  conditions,'' in \emph{Speech processing in the auditory system}.\hskip 1em
  plus 0.5em minus 0.4em\relax Springer, 2004, pp. 231--308.

\bibitem{bregman1994auditory}
A.~S. Bregman, \emph{Auditory scene analysis: The perceptual organization of
  sound}.\hskip 1em plus 0.5em minus 0.4em\relax MIT press, 1994.

\bibitem{darwin1995auditory}
C.~Darwin and R.~Carlyon, ``Auditory grouping, i in hearing. handbook of
  perception and cognition, bcj moore,'' 1995.

\bibitem{divenyi2004speech}
P.~Divenyi, \emph{Speech separation by humans and machines}.\hskip 1em plus
  0.5em minus 0.4em\relax Springer Science \& Business Media, 2004.

\bibitem{healy2017algorithm}
E.~W. Healy, M.~Delfarah, J.~L. Vasko, B.~L. Carter, and D.~Wang, ``An
  algorithm to increase intelligibility for hearing-impaired listeners in the
  presence of a competing talker,'' \emph{Journal of Acoustical Society of
  America}, vol. 141, no.~6, pp. 4230--4239, 2017.

\bibitem{van2017eeg}
S.~Van~Eyndhoven, T.~Francart, and A.~Bertrand, ``Eeg-informed attended speaker
  extraction from recorded speech mixtures with application in neuro-steered
  hearing prostheses,'' \emph{IEEE Trans. on Biomedical Engineering}, vol.~64,
  no.~5, pp. 1045--1056, 2017.

\bibitem{erdogan2017deep}
H.~Erdogan, J.~R. Hershey, S.~Watanabe, and J.~Le~Roux, ``Deep recurrent
  networks for separation and recognition of single-channel speech in
  nonstationary background audio,'' in \emph{New Era for Robust Speech
  Recognition}.\hskip 1em plus 0.5em minus 0.4em\relax Springer, 2017, pp.
  165--186.

\bibitem{rennie2010single}
S.~J. Rennie, J.~R. Hershey, and P.~A. Olsen, ``Single-channel multitalker
  speech recognition,'' \emph{IEEE Signal Processing Magazine}, vol.~27, no.~6,
  pp. 66--80, 2010.

\bibitem{weng2015deep}
C.~Weng, D.~Yu, M.~L. Seltzer, and J.~Droppo, ``Deep neural networks for
  single-channel multi-talker speech recognition,'' \emph{IEEE/ACM Transactions
  on Audio, Speech and Language Processing (TASLP)}, vol.~23, no.~10, pp.
  1670--1679, 2015.

\bibitem{sell2017multi}
G.~Sell and A.~McCree, ``Multi-speaker conversations, cross-talk, and
  diarization for speaker recognition,'' in \emph{2017 IEEE International
  Conference on Acoustics, Speech and Signal Processing (ICASSP)}.\hskip 1em
  plus 0.5em minus 0.4em\relax IEEE, 2017, pp. 5425--5429.

\bibitem{von2019all}
T.~von Neumann, K.~Kinoshita, M.~Delcroix, S.~Araki, T.~Nakatani, and
  R.~Haeb-Umbach, ``All-neural online source separation, counting, and
  diarization for meeting analysis,'' \emph{arXiv preprint arXiv:1902.07881},
  2019.

\bibitem{khorram2019jointly}
S.~Khorram, M.~McInnis, and E.~M. Provost, ``Jointly aligning and predicting
  continuous emotion annotations,'' \emph{IEEE Transactions on Affective
  Computing}, 2019.

\bibitem{gideon2017progressive}
J.~Gideon, S.~Khorram, Z.~Aldeneh, D.~Dimitriadis, and E.~M. Provost,
  ``Progressive neural networks for transfer learning in emotion recognition,''
  \emph{arXiv preprint arXiv:1706.03256}, 2017.

\bibitem{khalil2019robust}
M.~I. Khalil, N.~Mamun, and K.~Akter, ``A robust text dependent speaker
  identification using neural responses from the model of the auditory
  system,'' in \emph{International Conference on ECCE}.\hskip 1em plus 0.5em
  minus 0.4em\relax IEEE, 2019, pp. 1--4.

\bibitem{hansen2015speaker}
J.~H. Hansen and T.~Hasan, ``Speaker recognition by machines and humans: A
  tutorial review,'' \emph{IEEE Signal processing magazine}, vol.~32, no.~6,
  pp. 74--99, 2015.

\bibitem{ref1}
D.~Wang and G.Brown, ``Computational auditory scene analysis: Principles,
  algorithms, application.'' \emph{Wiley-IEEEPress}, 2006.

\bibitem{comon1994independent}
P.~Comon, ``Independent component analysis, a new concept?'' \emph{Signal
  processing}, vol.~36, no.~3, pp. 287--314, 1994.

\bibitem{roweis2001one}
S.~T. Roweis, ``One microphone source separation,'' in \emph{Advances in neural
  information processing systems}, 2001, pp. 793--799.

\bibitem{smaragdis2007convolutive}
P.~Smaragdis \emph{et~al.}, ``Convolutive speech bases and their application to
  supervised speech separation,'' \emph{IEEE Transactions on audio speech and
  language processing}, vol.~15, no.~1, p.~1, 2007.

\bibitem{yousefi2016supervised}
M.~Yousefi and M.~H. Savoji, ``Supervised speech enhancement using online
  group-sparse convolutive nmf,'' in \emph{International Symposium on
  Telecommunications (IST)}.\hskip 1em plus 0.5em minus 0.4em\relax IEEE, 2016,
  pp. 494--499.

\bibitem{wang2014training}
Y.~Wang, A.~Narayanan, and D.~Wang, ``On training targets for supervised speech
  separation,'' \emph{IEEE/ACM Transactions on Audio, Speech and Language
  Processing (TASLP)}, vol.~22, no.~12, pp. 1849--1858, 2014.

\bibitem{huang2015joint}
P.-S. Huang, M.~Kim, M.~Hasegawa-Johnson, and P.~Smaragdis, ``Joint
  optimization of masks and deep recurrent neural networks for monaural source
  separation,'' \emph{IEEE/ACM Trans. on Audio, Speech, and Lang. Proc.},
  vol.~23, no.~12, pp. 2136--2147, 2015.

\bibitem{zhang2016deep}
X.-L. Zhang and D.~Wang, ``A deep ensemble learning method for monaural speech
  separation,'' \emph{IEEE/ACM Transactions on Audio, Speech and Language
  Processing (TASLP)}, vol.~24, no.~5, pp. 967--977, 2016.

\bibitem{Mamun2019convolutional}
N.~Mamun, S.~Khorram, and J.~H.~L. Hansen, ``Convolutional neural network-based
  speech enhancement for cochlear implant recipients,'' in \emph{Proc.
  Interspeech}, 2019.

\bibitem{hershey2016deep}
J.~R. Hershey, Z.~Chen, J.~Le~Roux, and S.~Watanabe, ``Deep clustering:
  Discriminative embeddings for segmentation and separation,'' in
  \emph{Acoustics, Speech and Signal Processing (ICASSP), 2016 IEEE
  International Conference on}.\hskip 1em plus 0.5em minus 0.4em\relax IEEE,
  2016, pp. 31--35.

\bibitem{isik2016single}
Y.~Isik, J.~L. Roux, Z.~Chen, S.~Watanabe, and J.~R. Hershey, ``Single-channel
  multi-speaker separation using deep clustering,'' \emph{arXiv preprint
  arXiv:1607.02173}, 2016.

\bibitem{chen2017deep}
Z.~Chen, Y.~Luo, and N.~Mesgarani, ``Deep attractor network for
  single-microphone speaker separation,'' in \emph{Acoustics, Speech and Signal
  Processing (ICASSP), 2017 IEEE International Conference on}.\hskip 1em plus
  0.5em minus 0.4em\relax IEEE, 2017, pp. 246--250.

\bibitem{yu2017permutation}
D.~Yu, M.~Kolb{\ae}k, Z.-H. Tan, and J.~Jensen, ``Permutation invariant
  training of deep models for speaker-independent multi-talker speech
  separation,'' in \emph{Acoustics, Speech and Signal Processing (ICASSP), 2017
  IEEE International Conference on}.\hskip 1em plus 0.5em minus 0.4em\relax
  IEEE, 2017, pp. 241--245.

\bibitem{kolbaek2017multitalker}
M.~Kolb{\ae}k, D.~Yu, Z.-H. Tan, J.~Jensen, M.~Kolbaek, D.~Yu, Z.-H. Tan, and
  J.~Jensen, ``Multitalker speech separation with utterance-level permutation
  invariant training of deep recurrent neural networks,'' \emph{IEEE/ACM
  Transactions on Audio, Speech and Language Processing (TASLP)}, vol.~25,
  no.~10, pp. 1901--1913, 2017.

\bibitem{williamson2016complex}
D.~S. Williamson, Y.~Wang, and D.~Wang, ``Complex ratio masking for monaural
  speech separation,'' \emph{IEEE/ACM Transactions on Audio, Speech and
  Language Processing (TASLP)}, vol.~24, no.~3, pp. 483--492, 2016.

\bibitem{cuturi2017soft}
M.~Cuturi and M.~Blondel, ``Soft-dtw: a differentiable loss function for
  time-series,'' in \emph{Proceedings of the 34th International Conference on
  Machine Learning-Volume 70}.\hskip 1em plus 0.5em minus 0.4em\relax JMLR.
  org, 2017, pp. 894--903.

\bibitem{cooke2006audio}
M.~Cooke, J.~Barker, S.~Cunningham, and X.~Shao, ``An audio-visual corpus for
  speech perception and automatic speech recognition,'' \emph{The Journal of
  the Acoustical Society of America}, vol. 120, no.~5, pp. 2421--2424, 2006.

\bibitem{cooke2010monaural}
M.~Cooke, J.~R. Hershey, and S.~J. Rennie, ``Monaural speech separation and
  recognition challenge,'' \emph{Computer Speech \& Language}, vol.~24, no.~1,
  pp. 1--15, 2010.

\bibitem{shokouhi2017teager}
N.~Shokouhi and J.~H. Hansen, ``Teager--kaiser energy operators for overlapped
  speech detection,'' \emph{IEEE/ACM Transactions on Audio, Speech, and
  Language Processing}, vol.~25, no.~5, pp. 1035--1047, 2017.

\bibitem{yousefi2018assessing}
M.~Yousefi, N.~Shokouhi, and J.~Hansen, ``Assessing speaker engagement in
  2-person debates: Overlap detection in united states presidential debates,''
  in \emph{Proc. Interspeech}, 2018, pp. 2117--2121.

\bibitem{tu2015speech}
Y.-H. Tu, J.~Du, L.-R. Dai, and C.-H. Lee, ``Speech separation based on
  signal-noise-dependent deep neural networks for robust speech recognition,''
  in \emph{2015 IEEE International Conference on Acoustics, Speech and Signal
  Processing (ICASSP)}.\hskip 1em plus 0.5em minus 0.4em\relax IEEE, 2015, pp.
  61--65.

\bibitem{vincent2006performance}
E.~Vincent, R.~Gribonval, and C.~F{\'e}votte, ``Performance measurement in
  blind audio source separation,'' \emph{IEEE trans. on audio, speech, and
  lang. processing}, vol.~14, no.~4, pp. 1462--1469, 2006.

\bibitem{fox2007modeling}
B.~Fox, A.~Sabin, B.~Pardo, and A.~Zopf, ``Modeling perceptual similarity of
  audio signals for blind source separation evaluation,'' in
  \emph{International Conference on Independent Component Analysis and Signal
  Separation}.\hskip 1em plus 0.5em minus 0.4em\relax Springer, 2007, pp.
  454--461.

\bibitem{chen2017long}
J.~Chen and D.~Wang, ``Long short-term memory for speaker generalization in
  supervised speech separation,'' \emph{Journal of the Acoustical Society of
  America}, vol. 141, no.~6, pp. 4705--4714, 2017.

\end{thebibliography}
	
	
\end{document}